\documentstyle[aps,prb,epsf,psfig,twocolumn]{revtex}

\begin{document}
\draft \flushbottom \twocolumn[
\hsize\textwidth\columnwidth\hsize\csname
@twocolumnfalse\endcsname


\title{Rashba effect in 2D mesoscopic systems with transverse magnetic field}
\author{S. Bellucci $^1$ and P. Onorato $^1$ $^2$ \\}
\address{
        $^1$INFN, Laboratori Nazionali di Frascati,
        P.O. Box 13, 00044 Frascati, Italy. \\
        $^2$Dipartimento di Scienze Fisiche,
        Universit\`{a} degli Studi di Napoli ``Federico II'',
        Via Cintia, I-80126 Napoli, Italy.}
\date{\today}
\maketitle
\widetext
\date{\today}
\maketitle \widetext
\begin{abstract}
We present semiclassical and quantum mechanical results for the
effects of a strong magnetic field in Quantum Wires in the
presence of Rashba Spin Orbit coupling. Analytical and numerical
results show how the perturbation  acts in the presence of a
transverse magnetic field in the ballistic regime and we assume a
strong reduction of the backward scattering interaction which
could have some  consequences for the Tomonaga-Luttinger
transport. We analyze the spin texture due to the action of Spin
Orbit coupling and magnetic field often referring to the
semiclassical solutions that magnify the singular spin
polarization: results are obtained for free electrons in a
twodimensional electron gas and for electrons in a Quantum Wire.
 We propose the systems as
possible devices for the spin filtering at various regimes.
\end{abstract}


] \narrowtext
\tightenlines

\section{INTRODUCTION}

Recently the idea to use the electron spin
 in mesoscopic devices
has generated a lot of interest and the {\it "spintronic"} opens
new perspectives to semiconductor device technology and to quantum
computation. In  quantum computation the electron spin plays a
central role and offers unique possibilities in order to find new
mechanisms for information processing and information transmission
\cite{spintro,spintro2,wolf}. Datta and Das~\cite{Datta} describe
how the electrical field
 can be used to modulate the current:
essential for this mechanism is the field-dependent Spin Orbit(SO)
coupling.


The SO interaction has an essentially relativistic nature because
it stems directly from the quadratic in $v/c$ expansion of the
Dirac equation~\cite{Thankappan}. However, this perturbation can
give rise to a sensible  modification of a semiconductor band
structure \cite{Stormer,Nitta}.

The effects of an electric field on a moving electron have to be
analyzed starting from the SO hamiltonian
\begin{equation}
\hat H_{SO} = -\frac{\hbar}{(2M_0c)^2}\;{\bf E}({\bf R})
\left[\hat{{\bf \sigma}}\times \left\{\hat{\bf p}+\frac{e}{c}{\bf
A}({\bf R})\right\}\right]. \label{H_SO}
\end{equation}
Here $M_0$ is the free electron mass, $\hat{\bf p}$ is the
canonical momentum operator, $\hat{{\sigma}}$ are the Pauli
matrices, ${\bf E}({\bf R})$ is the electric field, ${\bf A}({\bf
R})$ is the vector potential, and ${\bf R}$ is the 3D position
vector.

The potential energy $V({\bf r})$ now  plays a central role
because if we are able to modulate it we can construct a current
modulator and a spin filter. In semiconductors heterostructures a
natural SO coupling is always present  because usually the
mesoscopic low dimensional devices are made at
 GaAs/AlGaAs interface obtained by
the MBE technology. In a triangular potential well at interface a
\emph{ two dimensional electron gas} (2DEG) is entrapped: the
interface electric field that accompanies the quantum-well
asymmetry~\cite{Kelly} is  directed along the normal to the
device plane hence it can be the source of the potential energy
$V({\bf r})$ in eq.(\ref{H_SO}). Because of the small dimension of
this well the electrons are fully confined in the growth ($z$)
direction, while their motion in the x-y plane is unrestricted.

Because it was first introduced by Rashba~\cite{Rashba}  the
mechanism of the SO interaction originating from the interface
field is known as Rashba effect. When  2DEG is confined in a
simple triangular well\cite{nota} along the $z$ direction by the
interface electric field $\overrightarrow{E}=(0,0,E_z)$  the
hamiltonian (\ref{H_SO}) becomes
\begin{equation}
\hat H_{SO} = \frac{\alpha}{\hbar}E_z M_0(\sigma_x V_y- \sigma_y V_x)
\label{H_SO2}
\end{equation}~
where ${\bf V}$ is the $velocity$ ${\bf V}={\bf p}+\frac{e}{c}{\bf
A}({\bf R})$ and corresponds to momentum if there is no external
magnetic field. Experimentally, in GaAs-AsGaAl interface, one
typically observes \cite{Nitta} values for $ \alpha E_z $ on the
order of $10^{-11}$ eV m. The SO coupling due to the electric
field in the $z$ direction is stronger than the Zeeman term of
interaction connected to a magnetic field acting on the system
because of the strong reduction of the effective electron mass
($m^*=0.068 m_0$). The Zeeman spin
 splitting term is $g^*\mu_B B$ where $g^*$ is the effective magnetic
 factor for
 electrons in this geometry (very low in GaAs) and $\mu_B$  is the
Bohr magneton
 with the bare mass.
So the mass renormalization reduces by $10$ order the SO coupling
and by $100$ the Zeeman spin splitting.

In this paper we discuss what happens when a strong transverse
magnetic field acts on a Quantum Wire (QW) in the presence of a
Rashba coupling. First we show a simple semi-classical approach
that suggests two possible mesoscopic devices for the spin
filtering, then we discuss the effects of a magnetic field on the
subbands structure of a Quantum Wire\cite{moroz,morozb} and show
the complex spin topology by analyzing the detailed spin textures
of the electron states\cite{governale}. The transport in QWs is
connected to three different regimes, the two ones at very low
temperatures correspond to the typical single electron tunneling
(Coulomb blockade) and to Ballistic Transport (where  the
Landauer-B\"uttiker formalism applies )\cite{BvH} while, when the
correlation is strong, the
Tomonaga-Luttinger\cite{moroz2,moroz3,egg} liquid regime dominates
and the broken simple $k,-k$ symmetry gives a subband structure
rather similar to the one of Single Wall Nanotubes. We suppose
that a possible effect of the magnetic field is the
back-scattering suppression due to the localization of the
different channels on the two different edges of the device.

\

\section{QUANTUM AND SEMI-CLASSICAL APPROACH}
We discuss some general effects of an electric field on a
classical  charge particle in two dimensions with an intrinsic
magnetic momentum $\bf \mu$. The  dynamics in the  presence of
external fields suggests some applications in the selection of
particles with polarized spin.

The simplest physical system consists of a  free particle moving
with initial momentum ${\bf p_0} = m{\bf v_0}$ in the plane. Using
eq.(\ref{H_SO2}) in absence of magnetic field the  total
hamiltonian is
\begin{eqnarray}
H=\frac{{\bf p}^2}{2m}+ \gamma E_z ({\bf \mu}\wedge{\bf p})
&\rightarrow& \nonumber \\ E=&\frac{m{ v_0}^2}{2}&+ \gamma E_z m
v_0 \mu sin(\vartheta)
\end{eqnarray} while the corresponding Hamilton
equations are $\dot{\bf p}=0$,
 $\dot{\bf x}={\bf v_0}(1+\gamma E_z m \mu  sin(\vartheta))$.
The difference between the unperturbed velocity ${\bf V}$  and the
real velocity $\dot{\bf x}$ plays a central role in the analysis
of the scattering at Wire-lead interface as we show in section IV.
A
beam of classical particles with fixed $\bf p$ is divided by an
electric field in different beams with different speed (depending
on the  angle $\vartheta$ between $\bf \mu$ and $\bf p$). This
could be the mechanism for a {\it Time Of Flight filter} which
uses
 the different time that particles with different
 magnetic moments employ to cross the same distance.

Besides this is the Rashba effect in quantum Mechanics. The Rashba
quantum effect  is due to the quantization of the magnetic
momentum ${\bf \mu}=\mu_0 {\bf \sigma}$ so that each beam with
definite momentum ${\bf p}=\hbar {\bf k}$ is split in two  beams
corresponding to the two opposite spin polarizations in the plane.
The linear momentum $ {\bf p}=\hbar \overrightarrow {k} $ commutes
with the free electron hamiltonian,
 thus
 we obtain the modified spectrum
in terms of $k=|\overrightarrow {k}|$
\begin{eqnarray}
\varepsilon_{f.e.}^\pm(k) = \frac{ {\hbar}^2 {(k \pm k_R)}^2}{2m}
-\frac{ {\hbar}^2 {k_R}^2}{2m}
\end{eqnarray}
where $k_R^0=\frac{m \alpha E_z }{\hbar}=\frac{p_R}{\hbar}$is
called  Rashba wavevector and  estimates the strength of the
splitting. The wavefunctions corresponding to the hamiltonian
$\hat{H}_0+ \hat{H}_{SO}$ are free waves in the orbital part while
the spinor is quantized along the direction orthogonal to the
motion. For ${\bf k}\parallel\hat{y}$ and ${\bf
E}\parallel\hat{z}$
$$
\phi_{k,s_x}(y)=\frac{e^{i k y}}{\sqrt{2 \pi L}}\chi^x_{s_x}.
$$
The spinors are in the $x-$base:
$\hat{\sigma}_x\chi^x_{\pm}=\pm (\hbar/2) \chi^x_{\pm}$ so that we define
$\hat{\sigma}_\pm=1/\sqrt{2}(\hat{\sigma}_y \pm i\hat{\sigma}_z)$.

Now we want show a simple semi-classical way to calculate the spin
polarization. We start from the classical solution of the
classical Hamilton (or Lagrange) equations, so we rewrite the
perturbative hamiltonian as follows (see appendix A):
\begin{equation}\label{ht}
  \hat{H}_{SO}= \frac{\alpha E_z M_0}{2} \left(\begin{array}{cc} V_y(t) &
-V_x(t) \cr -V_x(t)
&- V_y(t)
\end{array}\right),
\end{equation}
which then we can diagonalize in order to obtain an energy
correction and time dependent spinors. If we start from the simple
straight motion along the $\hat{y}$ axis we have ${\bf V}(t)=V_0
\hat{y}$ so that the splitting energy is $\Delta= \frac{\alpha E_z
M_0 V_0}{2}=\frac{\alpha E_z \hbar k}{2 }$. The spinors that
correspond to the eigenvalues of eq.(\ref{ht}) are the unperturbed
$\chi^x_\pm$.

An interesting case is the one of a particle in circular motion:
$V_x(t)=-\omega R \sin(\omega t)$ $V_y(t)=\omega R \cos(\omega t)$
where $\omega$ is the angular velocity and $R$ the orbit radius.
The energy splitting { reads} $\Delta= \frac{\alpha E_z   M_0
\omega R}{2}=\frac{\alpha E_z    \sqrt{M_0 \omega L_z} }{2 }$.
Now the spinors  are time dependent and the mean values of the
observables are  $\langle\hat{\sigma}_x (t)\rangle= \pm \sigma_0
\cos(\omega t)$ $\langle\hat{\sigma}_y (t)\rangle= \pm \sigma_0
\sin(\omega t)$.They  correspond to a spin precession at any time
orthogonal to the vector ${\bf V}(t)$. In Fig.(1) we show the spin
polarizations obtained from a semiclassical analysis.

\begin{figure}
\centering \epsfxsize= 9.40cm \epsfysize= 9.40cm
\epsfbox{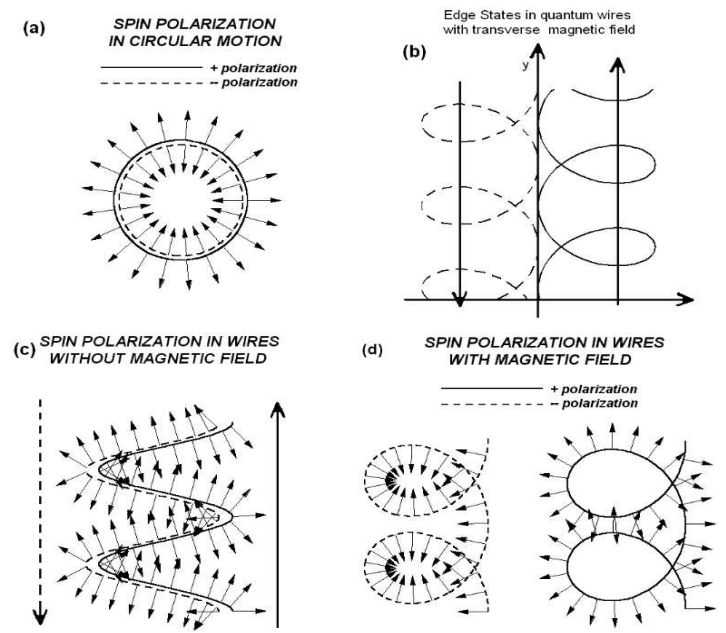}
 \caption{{ Semiclassical spin textures for
different orbits: a) spin polarization in circular motion;
b)classical edge states in a Wire under the effect of a transverse
magnetic field; c) spin polarization in a Wire without magnetic
field;  d)spin polarization for the edge
states.}}
\end{figure}

The case of an electron in a magnetic field directed along the
$\hat{z}$ axis is quite similar to the one discussed here, when we
consider the cyclotron frequency as the angular velocity
$\omega_c=\frac{eB}{mc}$.  Quantum Mechanical effects of the
Rashba hamiltonian on a free charge under the action of a magnetic
field are also discussed in the Appendix B.

\section{QUANTUM WIRE}

A Quantum Wire is usually defined by a parabolic confining
potential along one of the  directions in the plane:
$V(x)=\frac{M_0}{2}\omega_d^2 x^2$. We also consider
 a uniform magnetic field $B$ along the
$\hat{z}$ direction which allows a free choice in the gauge
determination. We choose the gauge so that the system has a
symmetry along the $\hat{y}$ direction: ${\bf A}=(0,Bx,0)$.
\begin{equation}\label{hnw}
H_{n.w.}  = M_0\frac{\bf V^2}{2}+\frac{M_0\omega_d^2}{2}x^2 +  \frac{\alpha}{\hbar}E_z M_0(\sigma_x V_y- \sigma_y V_x)
\end{equation}
where $M_0 V_y=p_y-eBx/(M_0 c)$ and $M_0 V_x=p_x$. This system has
also interesting limits: if we put  to zero the magnetic field we
have a simple narrow Wire \cite{governale} while
 if also $\omega_d$ vanishes we have a free  electron as in the case discussed in the previous section.
If  we let  only $\omega_d$ go to $0$ we obtain the electron in a
uniform magnetic field in transverse gauge and the spectrum has to
compare to the one found in the Appendix B  by using a different
gauge.

\subsection{Semi-classical solution}
In order to solve the hamiltonian in Quantum Wires we introduce
the total frequency $\omega_T=\sqrt{\omega_d^2+\omega_c^2}$ and
point out that    $p_y=V_y+eBx/(M_0c)$ commutes with the
hamiltonian
\begin{equation}\label{hnw2}
  H_{n.w.}  = \frac{\omega_d^2}{\omega_T^2}\frac{p_y^2}{2M_0}+\frac{p_x^2}{2M_0
  }+\frac{m \omega_T^2}{2}(x-x_0)^2 + H_{S.O.},
\end{equation}
where $x_0=\frac{\omega_c p_y}{\omega_T^2 M_0}$. The classical
Hamilton equations give us the orbital motion in the special case
of vanishing $\dot{x}(0)$
\begin{equation}\label{eqm}
  \begin{array}{ll} & x(t)=x_0+R \cos(\omega_T t) \cr
 & y(t)=V_d t - \frac{\omega_c}{\omega_T} R \sin(\omega_T t)+y(0)
\end{array}
\end{equation}
where the drift velocity $V_d=\frac{\omega_d^2 p_y}{\omega_T^2
M_0}$.
 We obtain $p_y=M_0(\dot{y}(0)+\omega_c x(0))$,  $y(0)=0$
and  $R=x(0)-x_0$ from the boundary conditions. The two different
motion along the Wire are localized on the two different edges as
we can argue from the introduction of $\pm p_y\rightarrow \pm
V_d$. These are also known in quantum mechanics as {\it edge
states}.

It's impossible to find an exact solution of the perturbed
hamiltonian as we can show by writing the usual semi-classical
approximation for the energy splitting: in this case the splitting
energy depends on time and oscillates between two values
$$
\Delta=\pm \frac{p_R}{2} \sqrt{\Sigma V^2+2 V_d \omega_c R
\cos(\omega_T t)-R^2 \omega_d^2 \cos^2(\omega_Tt)},
$$
where $\Sigma V^2=\left( V_d^2+\omega_T^2 R^2 \right)$. In Fig.(2)
we show how the perturbation affects the dispersion law for
vanishing magnetic field and strong SO coupling. Obviously there
are no deviations from "parabolicity". As we show in the next
subsection the subband deformation (Fig.(5)) is a second order
quantum effect due to the crossing between two nearest subbands
with opposite spin polarization. So in the classical case where no
subbands can exist the Rashba effect
 is as usual.
%
\begin{figure}
\centering \epsfxsize=8.40cm \epsfysize=8.40cm
\epsfbox{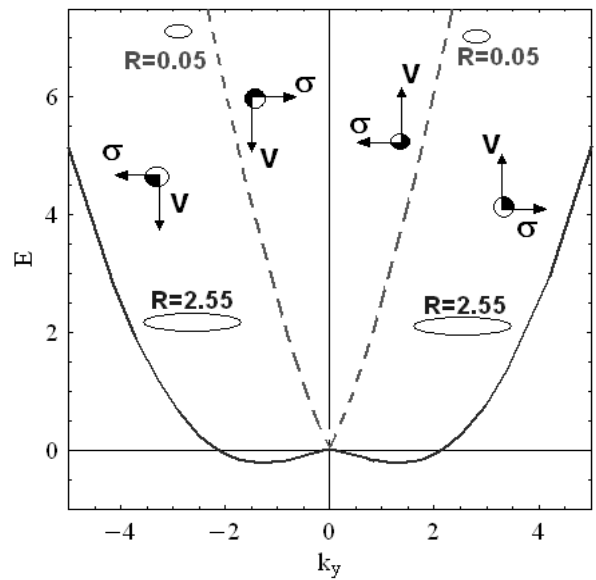}
\caption{{Semiclassical dispersion law obtained from the energy
after a mean value calculation. We show how the typical two split
parabolas have to be constructed starting from solutions with $\pm
\Delta$ but with different oscillations amplitude  ($R_+ \neq
R_-$). The semiclassical solutions give the dispersion law for a
spin forming a definite angle ($\pm\pi/2$) with the velocity,
while if we use the simple approach in the first section we have
two parabolas with definite spin along the $x$ axis.}}
\end{figure}
%
Now we need some prediction about the spin polarization induced by
the Rashba hamiltonian. Thus  we can use the general formula
eq.(\ref{sig}) in  Appendix A and calculate the spin polarization
depending on time. Obviously this result (showed in Fig.(1))
cannot be matched with Quantum Mechanical calculations and we have
to calculate the value of the spin component depending on
coordinates or on  momenta. So if we put $V_y=p_y /M_0$ and
$V_x^2=\omega^2(R^2-x^2)$ we have
$$
\langle\hat{\sigma}_x (x,p_y)\rangle= -
\frac{p_y}{\sqrt{p_y^2+M_0^2\omega^2(R^2-x^2)}}.
$$
\begin{figure}
\centering \epsfxsize= 8.40cm \epsfysize= 8.40cm
\epsfbox{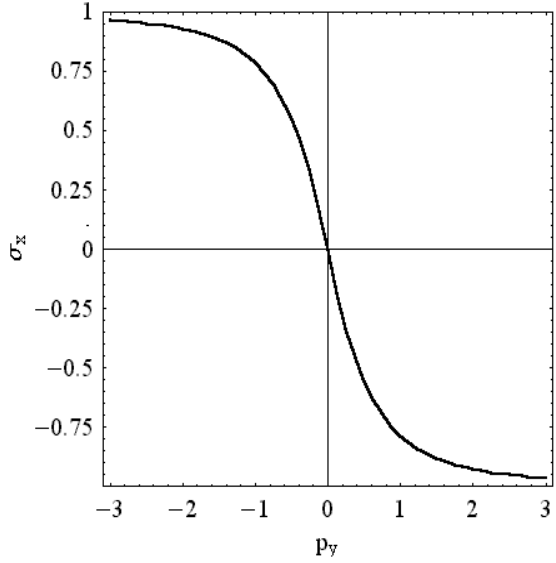}
 \caption{Expectation value of the
spin projection onto the $x$ direction depending on momentum
$p_y$.}
\end{figure}
\begin{figure}
\centering \epsfxsize= 9.40cm \epsfysize= 9.40cm
\epsfbox{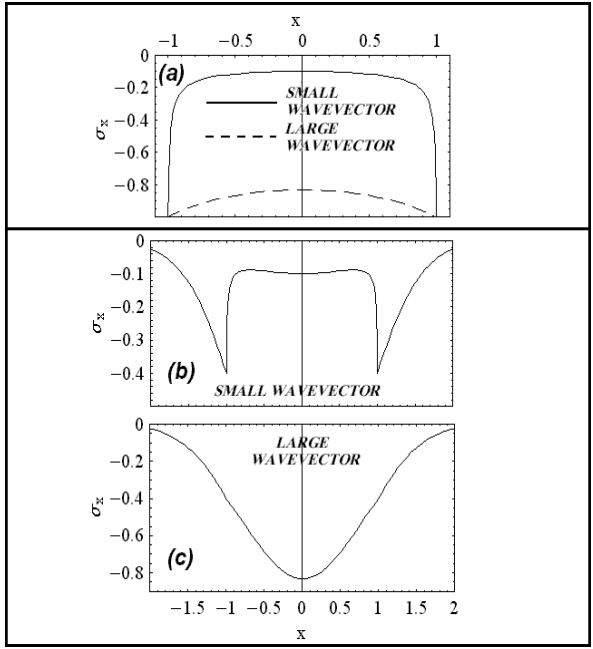}
 \caption{{Texture-like structure of the spin density across a
Quantum Wire, calculated within the semiclassical approach.
 a)Spin density along $x$ for large and small momenta.
b)and c) Results analogous to those by Governale and Zuliche
obtained by introducing a gaussian modulation of the spin density.
All  results were obtained without magnetic field.} }
\end{figure}
%
 The derivation of
$\langle\hat{\sigma}_x (p_y)\rangle$ is obtained by integrating in
$x$ between $-R$ and $R$. The results agree with Quantum
calculations by Governale and Zulicke \cite{governale} as we
 show in Fig.(3) that confirms conclusions reached in our
previous discussion about the spin structure. We can also
introduce a Gaussian correction which corresponds to the Quantum
harmonic oscillations simply by multiplying $\langle\hat{\sigma}_x
(x,p_y)\rangle$ by $\exp[-\frac{\omega M_0 x^2}{\hbar}]$ as showed
in Fig.(4.b-4.c) and compare Semiclassical results with
corresponding Quantum ones for small and large momenta
(Fig.(4.a-4.b-4.c)).

\subsection{Quantum Mechanical solution}

As we know from the semiclassical approach the perturbed Quantum
Wire has no exact solution and we can calculate its spectrum (and
related wavefunctions) by a numerical calculation or with simple
perturbation theory.

In order to introduce our result we can analyze the hamiltonian
eq.(\ref{H_SO2}) in the general case and separate the commuting
part (which gives an exact solution)
\begin{equation}\label{hd}
  \hat{H}_c=\frac{\alpha}{\hbar}E_z (p_y-M_0\omega_c x)
  \hat{\sigma}_x=\frac{2p_R}{\hbar M_0}\hat{\sigma}_x(p_y-M_0\omega_c x)
\end{equation}
and  the real perturbation
\begin{equation}\label{hnd}
  \hat{H}_n=\frac{\alpha}{\hbar}E_z p_x
  \hat{\sigma}_y=p_R \sqrt{\frac{2 \omega_T}{M_0 \hbar }}
(\hat{a}-\hat{a}^\dag)(\hat{\sigma}_++\hat{\sigma}_-)
\end{equation}

The first  order approximation has no contributions from
$\hat{H}_n$ and gives us the usual result with a linear
$k$-dependence that generates the Rashba subbands splitting
$$
H_{I^0}  = \frac{\omega_d^2}{\omega_T^2}\frac{(p_y-s
p_R)^2}{2M_0}-\frac{\omega_d^2}{\omega_T^2}\frac{p_R^2}{2M_0}+\frac{p_x^2}{2M_0}+\frac{M_0
\omega_T^2}{2}(x-\xi^s_0)^2,
$$
where $\xi^s_0=\frac{\omega_c (p_y-s p_R) }{\omega_T^2 M_0}$   and
$s=\pm 1$ correspond to $\chi_\pm$ spin eigenfunctions. Hence the
energies in the first  order approximation read
\begin{equation}\label{er}
 \varepsilon_{n,k,s} =\hbar
\omega_T(n+\frac{1}{2})+\frac{\omega_d^2}{\omega_T^2}\frac{\hbar^2}{2M_0}((k\pm
k_R)^2- k_R^2)
\end{equation}

From the first  order approximation we can conclude that 4-split
channels are present in the Quantum Wire corresponding to $\pm
p_y$ and $s=\pm1$. When a magnetic field is present the usual
energy and spin splitting\cite{moroz} correspond to the different
localization of the two channels and two different drift
velocities.

If we set the Fermi level at energy $E_F$ we find 4 Fermi momenta
 and related channels.

We can also observe that the distance between two different
channels increases with the magnetic field strength and has  an
upper bound at $\delta \xi=2\sqrt{\frac{2E_F}{ M_0}}$.
%
\begin{figure}
\centering \epsfxsize= 8.40cm \epsfysize= 8.40cm \epsfbox{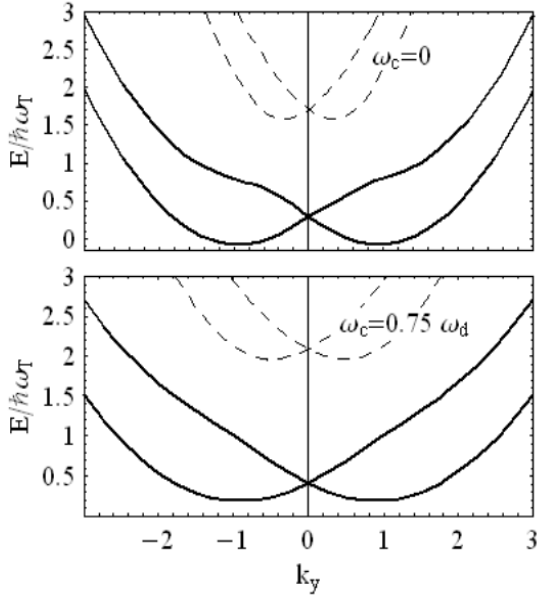}
\caption{{Lowest and first excited spin-split subbands of a
Quantum Wire, defined by a parabolic confining potential with
oscillator in a 2D electron system, with strong Rashba SO coupling
such that $p_R=1$. We compare the two different cases of a
vanishing magnetic field and a magnetic field of the same order of
the oscillator strength $\omega_c=.75 \omega_d$. As discussed in
the text the crossing momentum increases with the magnetic field
as $ k_y^c\propto(1+\frac{\omega_c^2}{\omega_d^2})^\frac{3}{2}$.
}}
\end{figure}
%
%
We cannot apply straightforwardly  perturbation theory because of
the crossing of the different subbands. From eq.(\ref{er}) we can
conclude that for $k_y^c=\pm\frac{M_0 \omega_T^3}{2 p_R
\omega_d^2}$ the two states connected by the hamiltonian
(\ref{hnd}) are degenerate and a strong deformation (the so called
avoided crossing) appears in the parabolic shape of the
unperturbed subband. A numerical implementation allows us to take
in account up to many subbands but the effect can also be seen in
the simple $3$ subband diagonalization. For the first subband we
can limit us to a simple $2\times2$ matrix which represents the
hamiltonian and we can give a simple analytical expression for the
wavefunction and the mean values of the spin components.

It is possible to obtain a good quantitative description of the
lowest spin-split subband by diagonalizing $H_{I^0}+H_n$ in a
truncated Hilbert space which is spanned by the lowest and
first-excited  parabolic subbands of the Hamiltonian as we discuss
in Appendix C. In Fig.(4) we show the deviation from parabolicity,
clearly visible near $\pm k_y^c$.
%
%
\begin{figure}
\centering \epsfxsize= 7.40cm \epsfysize= 7.40cm
\epsfbox{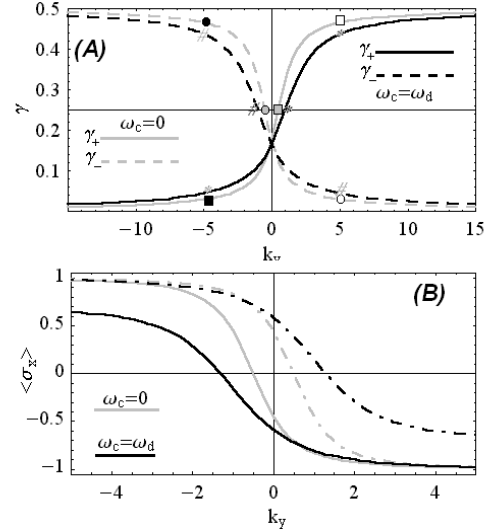}

\caption{{Spin structure of electron states in a quantum Wire with
strong SO coupling. A)$\gamma$ angle as defined in Appendix C as a
function of $k_y$ for both zero and a strong magnetic field. B)
Expectation value of the spin projection onto the $x$ direction
for electron states obtained in Fig.(5). Right-moving electrons
with large wave vectors asymptotically have parallel spin which is
opposite to that of the left-movers (dashing lines correspond to
the left parabola in Fig.(5) while  continuous ones correspond to
the right parabola).} }
\end{figure}
%
%
%
%
\begin{figure}
\centering \epsfxsize= 8.40cm \epsfysize= 8.40cm
\epsfbox{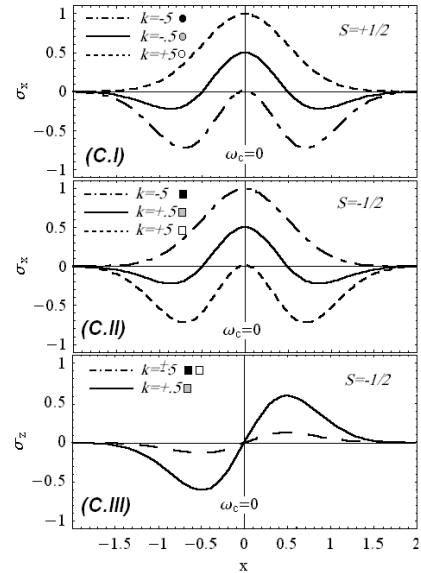}

\caption{{Spin structure of electron states in a quantum Wire with
strong SO coupling  without magnetic field.   Texture-like
structure of the spin density across the Quantum Wire, calculated
within the two-band model for the  states indicated by dots and
squares in Fig. (6.A). I)and II) ~Spatial variation of the $x$
components of the  spin density for the state with different wave
vectors and in  III) the $z$ component. } }
\end{figure}
{}
%
In Fig.(6) and (7) we show some graphs useful in order  to compare
our results with those obtained by Governale and Zuliche Fig.(6.B)
and Fig.(7) and we focus the effects of magnetic field in the
system. In Fig.(8) the effects of localization are clearly showed
in  three different regimes. Large $k_y$ of opposite signs
correspond to the single state limit ($i.e.$ $\gamma$ is $\pi/2$
or vanishes as showed in Fig.(6.A) ). The intermediate state
corresponding to the $k_y^c$ is a sum of the two states.
%
\begin{figure}
\centering \epsfxsize= 9.40cm \epsfysize= 9.40cm
\epsfbox{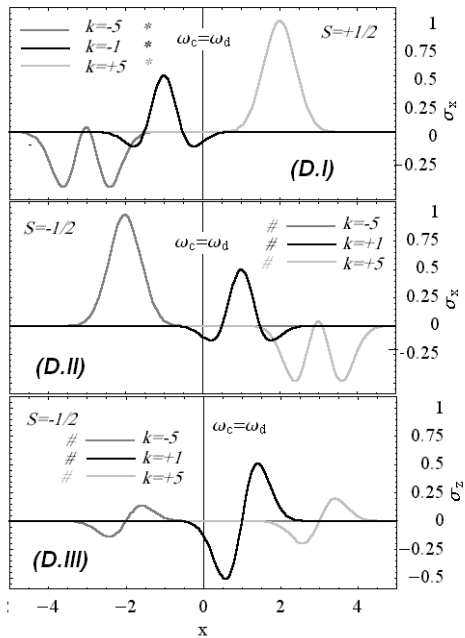}

\caption{{Spin structure of electron states in a quantum Wire with
strong SO coupling in the presence of a strong magnetic field
($\omega_c=\omega_d$). Texture-like structure of the spin density
across the Quantum Wire, calculated within the two-band model for
the  states indicated by $*$ and $\#$ in A). I)and II) ~Spatial
variation of the $x$ components of the spin density for the state
with different wave vectors and in III) the $z$ component. } }
\end{figure}
%

\section{Transport }

Now we  apply the Landauer-B\"uttiker
formalism\cite{landauer2,butt} and the
Tomonaga-Luttinger~\cite{TL,TLreviwew} model in order to discuss
spin--dependent transport. These models correspond to two
different regimes,  one where we suppose that there is  no
electron-electron interaction (Ballistic conductance) and the
opposite one where correlation effects due to the interaction
dominate.

\subsection{Non interacting electrons: Ballistic transport and B\"uttiker-Landauer formulation}
A general model for  near-equilibrium transport is due to the
Landauer\cite{landauer2} and B\"uttiker\cite{butt} contributions
that are condensed in the so called Landauer formula. This formula
expresses the conductance of a system at very low temperatures and
very small bias voltages calculated directly from the energy
spectrum by relating it to the number  of forward propagating
electron modes at a given Fermi energy
\begin{equation}
G \equiv G(\varepsilon_F) =
\frac{e^2}{h}\sum_{\sigma,n}\sum_{\sigma',n'}|t_{\sigma,\sigma',n,n'}(\varepsilon_F)|^2,
\label{G}
\end{equation}
 where $t_{\sigma,\sigma',n,n'}(\varepsilon_F)$
are  quantum mechanical transmission coefficients depending on the
subband index and the spin. If we suppose
$t_{\sigma,\sigma',n,n'}=\delta_{n,n'}\delta_{\sigma,\sigma'}$ we
can limit ourselves  to the number $M$ of forward propagating
electron modes at a given Fermi energy $\varepsilon_F$:
$G=(e^2/h)M(\varepsilon_F)$.

In a Quantum Wire, if the width   is smaller than the mean free
path  and comparable to the Fermi wavelength, the electrons move
in a perfectly coherent way throughout the device
 $i.e.$ the  current can travel adiabatically in any  state
without scattering into any other one~\cite{Baranger}, so we can
apply the ballistic theory.

Now we discuss the
 results obtained by Moroz and
Barnes\cite{moroz,morozb} simply by counting the modes showed in
Fig.(5). So we can  point out the effects of the transverse
magnetic field.

- In Fig.(5) we see a small {\it non-monotonic} portion (``bump'')
in  the energy curves $E(k_y)$ near $k^c_y$. Thus for the Fermi
energy $\sim E(k^c_y)$ we have $3$ propagating electron modes. Two
of them  have oppositely directed group velocities
$v_y=\hbar^{-1}(\partial E/\partial k_y)$ so the effects on the
conductance are damped. However the existence of such modes could
give rise to sharp peaks in the conductance $G(E_F)$. The magnetic
field shifts the peaks  and also attenuates the effect.

- A second manifestation of the $\alpha$-coupling is a shift of
the conductance quantization steps by $\delta E= \frac{\omega_d^2
p_R^2}{2\omega_T^2 M_0}$ to lower energies, in comparison with the
case of zero SO interaction. $\delta E$ vanishes for a large
magnetic field.

\

As discussed more recently by Governale and
Zulicke\cite{governale} the previous model could be not realistic
because  the SO coupling often vanishes in the contacts. In an
hybrid system  the transport  is strongly affected by the
scattering at Wire-lead interfaces due to the different nature of
electron states in the Wire and the leads.  The simplest
model~\cite{mire} is  the one obtained by attaching semi-infinite
leads with $p_{\text{R}}=0$ to the Wire where $p_{\text{R}}\ne 0$.
The transmission problem can be solved exactly by matching
appropriate {\it Ans\"atze\/} for wave functions in the Wire and
the leads. The usual condition for ensuring current conservation
has to be modified because the semiclassical  velocity for
electrons in the Quantum Wire\cite{mol,uz} is affected by the SO
coupling and the magnetic field:
\begin{equation}\label{velop}
v_y = \frac{\omega_d^2 p_y}{\omega_T^2 M_0}-\omega_c x(t)+
\frac{p_R \sigma_x(t)}{M_0}\quad .
\end{equation}
For a $1D$ Wire it is easy to calculate the usual transmission
value at interface\cite{mol} in eq.(\ref{t0}). This formula well
approximates the first order behaviour of the system but it does
not include the subband subband scattering which is strongly
enhanced by the magnetic field.


\begin{figure}
\centering \epsfxsize= 9.40cm \epsfysize= 4.70cm
\epsfbox{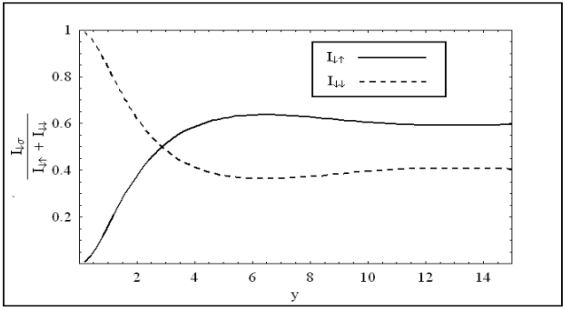}

 \caption{{Transport in a hybrid lead-Wire
system (a semi-infinite Wire ($y>0$) is attached to an ideal lead
($y<0$)) calculated using the Landauer--B\"uttiker formalism. In
the  Wire a strong Rashba SO coupling is present: we show the
conversion of incident spin-down current ($I_{\downarrow\sigma}$
denotes the spin--$\sigma$ current in the Wire when a
spin--$\downarrow$ current is injected from the lead). We suppose
$\gamma\approx \pi/3$, $p_R=.45$ and $k_0\approx 1.3$ in the usual
units ($M_0=1, \omega_d=1, \hbar=1$). }}
\end{figure}
{}
If the SO coupling is small ($p_R<<\sqrt{\hbar M_0 \omega}$) we
can limit ourselves to the  analysis of the spin
precession\cite{Datta,mire} due to the Rashba hamiltonian. In this
case we can limit ourselves to  the first order perturbation with
$t^0_\sigma$ and $H_{I^0}$. We conclude that a modulation of the
transmitted current at drain appears when we inject a spin
polarized current from the emitter to the Wire.

When the SO Coupling is very strong we have to turn to the two (or
more) band model discussed in the previous section. In this case a
current conversion is enabled by scattering into evanescent modes
of the Wire, so that we have different behaviours in the current
polarization
 close to the interfaces in
the Wire  and very far from it. We discuss all that in appendix D
and we show the results  in Fig.(9). In this figure we show the
conversion of spin down incident current depending on the distance
from the Wire-lead interface.
\begin{figure}
\centering \epsfxsize= 9.40cm \epsfysize= 4.70cm
\epsfbox{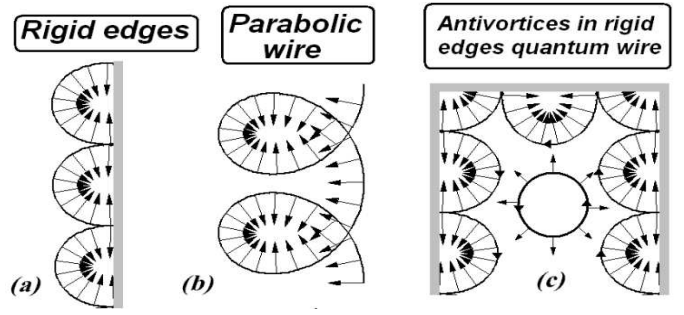}

 \caption{ In a) and b) we compare a
rigid wells Wire and a parabolic one and give a heuristic
explanation of the spin flip at the interface. In c) we show the
mechanism for the current antivortices and the related spin
polarization in the classical case. }
\end{figure}

\subsubsection{Current and Spin vortices in Quantum Wires}

When we consider the scattering of the current from the Wire-Lead
interface (see Appendix D) we can have a very complex spin
polarization related to the current.
 The presence of current  antivortices in the Wire when the reflected current
does not vanish is known (as the presence of vortices connected to
the transmitted one\cite{MRZ}) and can be easily explained in the
classical case of a Wire with rigid wells Fig.(10). The
antivortices are near the $y$ axis if the transmission is very
large, as we show in Fig.(11 and 12).


\begin{figure}
\centering \epsfxsize= 7.40cm \epsfysize= 7.40cm
\epsfbox{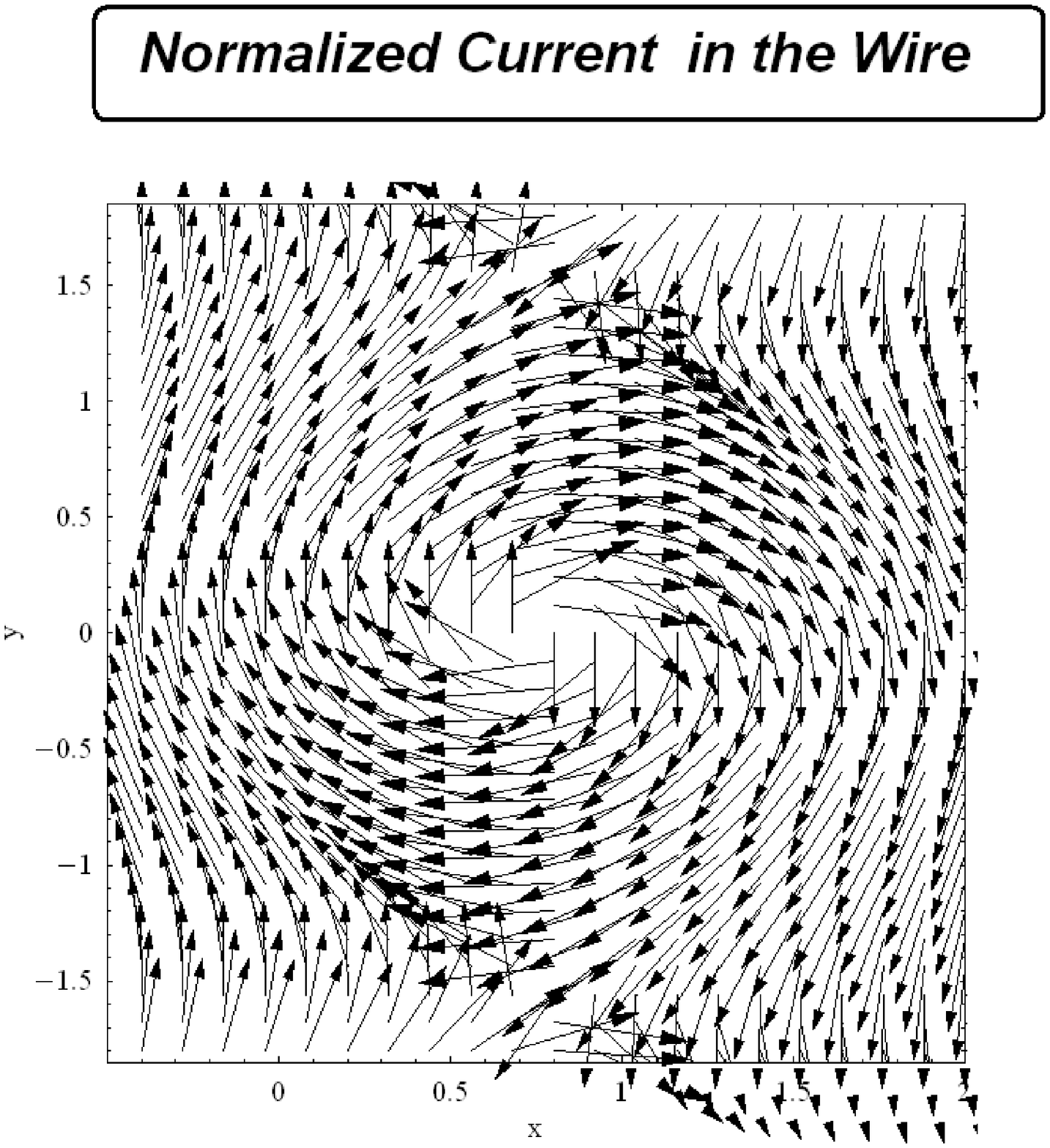}

 \caption{{  We show the vector
field of the current (eq.(\ref{j})) renormalized with respect to
its modulus around an antivortex. The antivortices go toward the
$y$ axis when the energy increases.  }}
\end{figure}


\begin{figure}
\centering \epsfxsize= 7.40cm \epsfysize= 7.40cm
\epsfbox{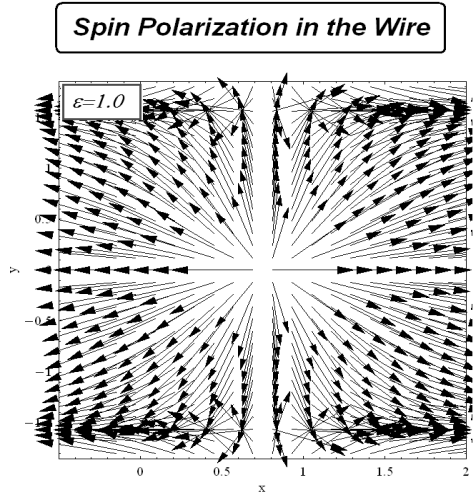}

 \caption{{
 We show the in plane spin texture near
the antivortices. We can compare this texture with the one showed
in Fig(10) for a rigid well wire and with the one in Fig.(1.a) }}
\end{figure}


The quantum mechanical results, Fig.(11 and 12), confirm the
classical picture giving a very singular spin texture all around
the antivortices.

\subsection{Interacting electrons: effective theory and Tomonaga-Luttinger model}

As we discussed in the previous sections the strong effect of a
magnetic field consists in the localization of the left and right
going electrons on the opposite sides of the Wire. This effect
correspond to a sort of $1$-dimensionality suppression and reveals
the $2$-dimensional behaviour of the Wire.  The magnetic field
competes with the lateral confining potential in order to
determine the exact dimension of the system. This could open new
perspectives in the use of dimensional crossover\cite{14,24,34,44} in
order to analyze the Fermi-non Fermi liquid behaviour of
interacting electrons in the Wire.

Next we refer to the usual approach to the Luttinger liquid in
Quantum Wires under a strong Rashba SO
coupling\cite{moroz2,governale,h}; this analysis  is not in
general correct for  $1$D systems but just for the limited class
of semiconducting Quantum Wires as discussed in Ref.\cite{egg}
where the Carbon Nanotube case is analyzed.

The effective low-energy description of an interacting quantum
Wire starts from Tomonaga-Luttinger models\cite{TL}, so we have to
linearize the single-electron energy spectrum close to the four
Fermi points: we obtain two different  electron velocities
 $v_{1,2}=\partial E(p_{1,2})/\partial p$  derived from the
dispersion law in Fig.(5). Following  Ref.\cite{moroz2} we neglect
 Umklapp scattering, a legitimate assumption  if the  energy
bands are far from being half-filled~\cite{TLreviwew}, which is
exactly the case, e.g., in Quantum Wires patterned in
semiconductor heterostructures~\cite{Kelly}.

In a future paper we intend to investigate the magnetic field
 effects on the Luttinger model  parameters with some details.
Now we just suppose that the magnetic field  could improve the
validity of the elimination of backward scattering especially when
$\omega_c\approx \omega_d$ because of the localization of the edge
states. In fact the distance between two electron with opposite
momenta ($k_y$,$-k_y$) has a mean value increased by magnetic
field as $\delta r \propto \frac{\omega_c}{\omega_T}k_y$. In the
same way  also forward scattering between electrons on the
opposite branches (the usual $g_2$ coupling) is affected by the
localization. So we can assume that a sort of chiral Luttinger
liquid behaviour (where the non vanishing interaction parameter is
just $g_4$) takes place in the Wire when the   magnetic field is
strong without the spin-charge separation.

 So we can conclude as in
Ref.\cite{moroz2} that the SO coupling mixes together the charge
and spin excitations  and thereby  destroys the usual spin-charge
separation  in the Tomonaga-Luttinger liquids.

\
\begin{center}
  {\bf Acknowledgement}
\end{center}

\noindent This work is partly supported by the Italian Research
Ministry, National Interest Program "Effetti di spin, interazione
e proprieta` di trasporto in sistemi elettronici fortemente
interagenti a bassa dimensionalit\`a".

\
\appendix
\section{General solution of the semi-classical problem and observables}
\label{a1}
 Let us introduce the three spin operators
$\hat{\sigma}_i$ in the $x-$based representation:
$$
\hat{\sigma}_x= \frac{\hbar}{2} \left(\begin{array}{cc} 1 & 0 \cr0
& -1
\end{array}\right),
 \hat{\sigma}_y= \frac{\hbar}{2} \left(\begin{array}{cc} 0 &
1 \cr 1 & 0
\end{array}\right),
 \hat{\sigma}_z= \frac{\hbar}{2} \left(\begin{array}{cc} 0 &
i \cr-i & 0
\end{array}\right)
$$
The Rashba hamiltonian in semi-classical approach is showed in
eq.(\ref{ht}) and can be easily diagonalized giving the result:
\begin{equation}\label{fa1}
  \Delta\varepsilon_\pm=\pm\ \frac{p_R}{2
\hbar^2}\sqrt{V_x(t)^2+V_y(t)^2}.
\end{equation}
which corresponds to an energy shift of the  first order in $p_R$.
The knowledge of the eigenvectors allows the calculation of the
spin components
\begin{eqnarray}\label{sig}
  \begin{array}{c}
    \langle\hat{\sigma}_x (t)\rangle= \pm
\frac{V_y(t)}{|\sqrt{V_x(t)^2+V_y(t)^2}|} \\
   \langle\hat{\sigma}_y (t)\rangle= \mp
\frac{V_x(t)}{|\sqrt{V_x(t)^2+V_y(t)^2}|} \
  \end{array}.
\end{eqnarray}
In order to demonstrate that the spin polarization rotates we can
solve the Hamilton equations to the  first perturbative order. We
start from the hamiltonian (\ref{H_SO2}) and from the unperturbed
orbits: $x(t),y(t),V_x(t),V_y(t)$. Thus  we introduce the hamilton
equation at first order in $p_R$:
\begin{eqnarray}\label{sp}
  \begin{array}{l}
   \dot{\sigma}_x =\frac{p_R}{2}V_x {\sigma}_z  \\
 \dot{\sigma}_y  =\frac{p_R}{2}V_y {\sigma}_z \\
  \dot{\sigma}_z=\frac{p_R}{2}\left(V_y {\sigma}_y + V_x {\sigma}_x\right)
  \end{array}
\end{eqnarray}
It is simple to verify  the general propriety: $[\dot{\bf
\sigma}\times{\bf V}]_{z}=0$.

In order to solve  eq.(\ref{sp}) we can introduce a new spin
vector rather similar to a polar one. So we introduce
\begin{equation}\label{slo}
\hat{\sigma}_\parallel= \frac{V_x
\hat{\sigma}_x+V_y\hat{\sigma}_y}{\sqrt{V_x^2+V_y^2}} ,
 \hat{\sigma}_\perp= \frac{V_x
\hat{\sigma}_y- V_y\hat{\sigma}_x}{\sqrt{V_x^2+V_y^2}}
\end{equation}
with the usual commutation rule
$[\hat{\sigma}_\parallel,\hat{\sigma}_\perp]=i\hbar\hat{\sigma}_z$.
So if we suppose that the orbits are unperturbed we obtain for the
eq.(\ref{sp}):
\begin{eqnarray}\label{s2p}
  \begin{array}{l}
   \dot{\sigma}_\perp =0  \\
 \dot{\sigma}_\parallel  =\frac{p_R}{2}\sqrt{V_x^2+V_y^2} {\sigma}_z \\
  \dot{\sigma}_z=-\frac{p_R}{2}\sqrt{V_x^2+V_y^2}{\sigma}_\parallel
  \end{array}
\end{eqnarray}
These equations have a simple solutions if $\sqrt{V_x^2+V_y^2}$
does not depend on time, with two oscillating components of the
spin vector with frequency
$\omega_p=\frac{p_R}{2}\sqrt{V_x^2+V_y^2}$. However the first
order correction to the energy correspond to that one in
eq.(\ref{fa1}).
\section{Rashba effect in the symmetric gauge}
\label{a2}

In this section we show the effect of a transverse magnetic field
on a free electron when SO perturbation is also present. The
difference with the calculation of Section III in the limit of
vanishing potential ($V(x)$) is the different gauge that we
choose. Thanks to this choice we can exactly solve the quantum
problem. The symmetric gauge is ${\bf A}=(-By/2,Bx/2,0)$ so that
we can define:
\begin{eqnarray}\label{si}
  \begin{array}{ll}
  V_x=p_x+\frac{e B y}{2 M_0 c} & V_y=p_y-\frac{e B x}{2 M_0 c} \\
 V_+=\frac{1}{\sqrt{2}}(V_x+iV_y) & V_-=\frac{1}{\sqrt{2}}(V_x-iV_y)\\
 \hat{\sigma}_+^z=\frac{1}{\sqrt{2}}(\hat{\sigma}_x+ i
 \hat{\sigma}_y) & \hat{\sigma}_-^z=\frac{1}{\sqrt{2}}(\hat{\sigma}_x- i
 \hat{\sigma}_y)
  \end{array}.
\end{eqnarray}
The hamiltonian reads
$$
  H_0+H_{SO}=\frac{M_0}{2} (V_-V_++ V_+V_-)+
  i \frac{p_R}{ \hbar}(V_-\sigma_+ -V_+\sigma_-).
$$
From the commutation rules $[V_+,V_-]=\frac{\hbar \omega_c}{2
M_0}$ we can deduce that $V_\pm$ correspond to $a_-$ an $a_-^\dag$
operators in the 2D isotropic harmonic oscillator. Hence we can
rewrite the previous hamiltonian as follows:
\begin{eqnarray}\label{hmf}
  H_0+H_{SO}=&\hbar \omega_c&(a_-^\dag a_- +\frac{1}{2} )+\nonumber \\&i \frac{p_R}{\hbar}&\sqrt{\frac{\hbar \omega_c}{
M_0}}(a_-^\dag\sigma_+ -a_-\sigma_-)
\end{eqnarray}
where the angular momentum is given by $l_z=m=n_+-n_-$ and $n_-$
labels the Landau levels with energy $(n_- +\frac{1}{2} )\hbar
\omega_c$.
 The
spin coupling term in eq.(\ref{hmf}) appears as a perturbation but
can be also exactly diagonalized, because it connects just two
nearest Landau Levels with the same value of
$j_z=m+s_z=n_+-n_-+s_z$. Therefore, also $n_+$ is an invariant.

The unperturbed wave-functions are
$$
\psi_{k,m}^{\rm L}(\vec{r})={\rm(const.)}\exp({\rm
i}m\varphi)r^{|m|} \exp(-B r^2) L_k^{|m|}(2Br^2)\label{Landau}
$$
where $\vec{r}=(r\,\cos\varphi,r\,\sin \varphi)$ and $L_k^{|m|}$
is an associated Laguerre polynomial and $
k=n_--\frac{1}{2}(|m|-m)$. For sake of simplicity we choose $m>0$
. So we can start from the unperturbed eigenfunctions
$\psi_{n_-,m}^{\rm L}\chi_\uparrow^z=|n_-,n_+,\uparrow\rangle$ and
the corresponding $j_z=(n_+-n_-)+1/2$ states
$|n_--1,n_+,\downarrow\rangle$.

In this $2\times2$ subspace the hamiltonian eq.(\ref{hmf}) can be
easily diagonalized with eigenvalues
$$
\varepsilon_{n_-,\pm}^{j_z,n_+}=n_- \hbar \omega_c \pm
\sqrt{\left(\frac{\hbar \omega_c}{2}\right)^2+\frac{p_R^2
\omega_c}{4 M_0 \hbar}n_-}
$$
while the lowest Landau level with spin up is  not split. Finally we
can introduce the perturbed wave functions as:
\begin{eqnarray}\label{no}
  |{n_-,\pm}\rangle_{j_z,n_+}=
\cos(\gamma_\pm)|n_--1,n_+,\downarrow\rangle+i
\sin(\gamma_\pm)|n_,n_+,\uparrow\rangle \nonumber
\end{eqnarray}
where
$$
\tan(\gamma_\pm)=-\frac{\frac{\hbar \omega_c}{2} \pm
\sqrt{\left(\frac{\hbar \omega_c}{2}\right)^2+\frac{p_R^2
\omega_c}{4 M_0 \hbar}n_-}}{\sqrt{\frac{p_R^2 \omega_c}{4 M_0
\hbar}n_-}}
$$
Thus we can calculate the mean values of the spin components
\begin{eqnarray}\label{no2}
 \langle \sigma_x(r,\varphi)
 \rangle= F(r) \sin(2\gamma_\pm) \cos(\varphi)\\
\langle \sigma_y (r,\varphi)
 \rangle= F(r) \sin(2\gamma_\pm) \sin(\varphi)
\end{eqnarray}
where $F(r)$ is a radial function depending on $n_-$ and $n_+$
easily obtained from the unperturbed wavefunctions.

\section{Two subbands analytic solution for the lowest level in a Quantum Wire}

We start from the spectrum in eq.(\ref{er}) and we consider the
effects on the ground states (GS) with opposite spin of the
hamiltonian $\hat{H}_n$ (\ref{hnd}). If we chose the GS with spin
up ($v_0=|0,k_y,\uparrow\rangle$) the effect of $\hat{H}_n$ is to
increase the subband label and decrease the spin
($v_1=|1,k_y,\downarrow\rangle$). In the matrix representation
the total hamiltonian in the subspace ($\{v_0,v_1\}$) becomes
\begin{eqnarray}\label{mh}
H=\left(\hbar \omega_T +\frac{\omega_d^2}{\omega_T^2}\frac{\hbar^2}{2M_0} k_y^2\right)\left(\begin{array}{cc}
    1&0 \\
   0&1
  \end{array}
\right)+
  \left(\begin{array}{cc}
    a & i b \\
  -i b& a  \end{array}\right)
\end{eqnarray}
where
$$
a= \left( \frac{\hbar \omega_T}{2}+
\frac{\omega_d^2}{\omega_T^2}\frac{\hbar k_y p_R}{M_0} \right),
\quad  \qquad b=p_R \sqrt{\frac{2 \omega_T}{M_0 \hbar }},
$$
so that we can easily diagonalize and obtain
$$
\Delta \varepsilon =\sqrt{a^2+b^2}, \quad \qquad
\tan(\gamma_\pm)=\frac{a\mp\Delta \varepsilon}{b}.
$$
So the new eigenvectors have the form:
$$
\varphi
= e^{i k_y y}\left(\cos(\gamma_\pm) u_{0,k_y,\uparrow}(x)\chi_\uparrow^x+ i \sin(\gamma_\pm)u_{1,k_y,\downarrow}(x)\chi_\downarrow^x\right)
 $$
 where $u_{n,k_y,\uparrow}(x)=\exp{(x-\xi_0^s)^2} h_n(x-\xi_0^s)$ with $\xi_0^s$ depending on $k_y$  so that
 $$
\langle\hat{\sigma}_x (x,k_y)\rangle= \frac{\cos(\gamma_\pm)^2
u_{0,k_y,\uparrow}(x)^2-\sin(\gamma_\pm)^2u_{1,k_y,\downarrow}(x)^2}{\cos(\gamma_\pm)^2
u_{0,k_y,\uparrow}(x)^2+\sin(\gamma_\pm)^2u_{1,k_y,\downarrow}(x)^2}.
$$
\section{Simple analytical solution for the lead-Wire scattering}

In order to apply the Landauer formula we need a method to
determinate the transmission and reflection coefficients for a
given energy. So we have to solve the Shr\"odinger equation near
the lead-Wire interface ($y=0$).

The first step is the solution of the first order problem  for the
lowest subband ($n=0$). Starting from eq.(\ref{er}), we fix the
Fermi energy and  find the wavevectors in the Wire $k_w$ and the
lead $k_L$ depending on the spin label $s_x$
$$
k_w^s(\varepsilon_F)=s k_R\pm \sqrt{k_R^2+k_0^2} \qquad
k_L(\varepsilon_F)=\pm k_0
$$
where
$$
k_0^2=2 M_0
\frac{\omega_T^2}{\omega_d^2}\left(\varepsilon_F-\frac{\hbar
\omega_T}{2}\right).
$$
Next, we inject the electron with spin $s_z=1/2$ corresponding to
the spinor $\frac{1}{\sqrt{2}}\left(\begin{array}{c}
    i \\
   1
  \end{array}
\right)$
\begin{eqnarray}
 \frac{u_0^{}(x)}{\sqrt{2}}e^{i k_0
y}\left(\begin{array}{c}
    i \\
   1
  \end{array}
\right)&+& r_\uparrow  u_0(x)e^{-i k_0 y}\left(\begin{array}{c}
    1 \\
   0
  \end{array}
\right) \nonumber \\ &+& r_\downarrow u_0(x)e^{-i k_0
y}\left(\begin{array}{c}
    0 \\
   1
  \end{array} \nonumber
\right) \end{eqnarray}
$$+ t_\uparrow  u_0^{k,\uparrow}(x)e^{i k_w^\uparrow
y}\left(\begin{array}{c}
    1 \\
   0
  \end{array}
\right)+t_\downarrow u_0^{k,\downarrow}(x)e^{i k_w^\downarrow
y}\left(\begin{array}{c}
    0 \\
   1
  \end{array}
\right).
$$
Then we introduce the conditions for the continuity of the wave
function and the current flux ($-i\partial_y-p_R/m
\hat{\sigma}_x$) at the interface . The simple case of vanishing
magnetic field can be solved starting from the equivalence
$u_0(x)=u_0^{k,s}(x)$ and implies two identical systems of
equations for $\uparrow$ and $\downarrow$ spins
\begin{equation}\label{t0}
  t^0_\uparrow=i t^0_\downarrow = i\frac{\sqrt{2}
  k_0}{k_0+\sqrt{k_0^2+k_R^2}}.
\end{equation}
From these coefficients we can evaluate the mean value of
$\hat{\sigma}_z$ and the probability of detecting a spin up (down)
electron at the collector $P_{s_z=1}(y=L)$ at interface
\begin{eqnarray}
\langle \hat{\sigma}_z\rangle&=&\frac{\cos(2k_RL)}{2} \nonumber\\
P_{s_z=1}\propto\cos(k_RL)^2 \quad & & \quad
P_{s_z=-1}\propto\sin(k_RL)^2.\nonumber
\end{eqnarray}
This is the basis of Datta and Das device for spin filtering.

The strong Rashba effect for vanishing magnetic field is just a
little more difficult: we have to introduce the SO called
evanescent states from the higher subbands while the transmitted
states in the Wire are not any  more pure $u_0(x)$ but the more
complex function $\varphi$. If we are near the $k_y^c$ energy just
the $\downarrow$ eigenfunction is affected by the second order
perturbation while the $\uparrow$ one is as in the previous case.
Here we do not discuss the more complex case of non monotonicity
discussed in the text so we have to modify the wavefunction in the
Wire in a very simple manner
\begin{eqnarray}
t_\uparrow  u_0(x)e^{i k_w^\uparrow y}\left(\begin{array}{c}
    1 \\
   0
  \end{array}
\right)+
 \tau_\uparrow  u_1(x)e^{i k_Ry -q
y}\left(\begin{array}{c}
    1 \nonumber \\
   0
  \end{array}
\right)+ \\ + t_\downarrow \left(\cos(\gamma) u_0(x)e^{i
k_w^\downarrow y}\left(\begin{array}{c}
    0 \\
   1
  \end{array}
\right)+i \sin(\gamma)u_1(x)e^{i k_w^\downarrow
y}\left(\begin{array}{c}
    1 \\
   0
  \end{array}
\right) \right) \nonumber \end{eqnarray}
 and also add a reflected evanescent channel in the lead
 $$
 \rho_\uparrow u_1(x)e^{\chi_0 y}
 $$
 where $\chi_0^2=2 M_0
\frac{\omega_T^2}{\omega_d^2}\left(\frac{3\hbar
\omega_T}{2}-\varepsilon_F\right)$ and $q^2=\chi_0^2+k_R^2$.

If we remember that $\langle u_0|u_1\rangle =0$ we obtain three
systems of equations corresponding to  $\left(\begin{array}{c}
    1 \\
   0
  \end{array}\right)u_0$,
  $\left(\begin{array}{c}
    1 \\
   0
  \end{array}\right)u_1$,
  $\left(\begin{array}{c}
    0 \\
   1
  \end{array}\right)u_0$
  easily solvable in the coefficients. We also recall that
  no spin-polarized current is generated in the leads.
  So we can put $\rho_\uparrow=0$ and obtain
\begin{equation}\label{t1}
  t^I_\uparrow=t^0_\uparrow;\quad  t^I_\downarrow = i\frac{t^0_\uparrow}{\cos(\gamma)}
  \quad  \tau^I_\uparrow = \tan(\gamma)t^0_\uparrow
\end{equation}
Now we can use the usual B\"uttiker-Landauer
formulation\cite{landauer2,butt}.   So we confirm the results
about the spin polarization in Ref.\cite{governale} as we show in
Fig.(9). We have a  generated spin-up current from an incident
spin-down one
$$
I_{\downarrow \uparrow}=k_\Delta\left(1-e^{-q y}\cos\left[k_\Delta
y \right]\right)\tan^2(\gamma)|t^0_\uparrow|^2
$$
where $k_\Sigma=\sqrt{k_R^2+k_0^2}$ and $k_\Delta=k_\Sigma-2 k_R$.
The damped oscillations in the current vanish when the distance
from the interface increases and we have a polarized current.

\subsubsection{Reflection by a Wire-lead interface in the presence of a magnetic field}

We can start from the spinless case of one electron reflected by
an interface. We have to remember that in this case the
wavefunctions are localized in the opposite edges of the Wire. The
simple solutions of the previous section are now not the exact
ones but we also use the simple $1D$ approach that  could yield a
good approximation. However is clear that, also in absence of
spin, evanescent states have to be included in order to have an
exact calculation of the transmission coefficients.

A fundamental step in our analysis is the definition of a correct
{\it Ans\"atz} for the spin conservation in the scattering. If we
compare the spin polarization in the edge states in a parabolic
confined Wire Fig.(1.d) with the edge states reflected by rigid
edges we can conclude that in the reflection from a rigid well the
spin polarization is not conserved. We can show that the wedge
product between velocity and spin is preserved as we show in
Fig.(10). We suppose that the conserved quantity be $(\vec{\bf
\sigma}\times\vec{\bf v})_z$ {\it i.e.} we have in the reflection
$$
\vec{\bf v}\rightarrow-\vec{\bf v}\Rightarrow\vec{\bf
\sigma}_p\rightarrow-\vec{\bf \sigma}_p \quad
\sigma_z\rightarrow\sigma_z
$$
where $\vec{\bf \sigma}_p$ is the spin in the $x-y$ plane.

Now we calculate the one channel model corresponding to the
incident $|k,\uparrow\rangle$ state, the reflected
$r_\downarrow|-k,\downarrow\rangle$ and the transmitted
$t_\uparrow|k_0,\uparrow\rangle$ state where $k=k_0+k_R$. The
reflection coefficient reads
$$r_\downarrow=\frac{k_R}{2 k_0+k_R}=e^{-2 \Delta_R}.$$
Thus we can express the wavefunction in the Quantum Wire to first
order as follows:
\begin{eqnarray}
\psi(x,y)&=&N
e^{-\frac{(x^2+\xi^2)}{2}-\Delta_R}\left(\begin{array}{c}
    e^{x\xi+\Delta_R}e^{i k y
} \\
   e^{-x\xi-\Delta_R}e^{-i k y
}
  \end{array}\right)\nonumber \\
&=& f(x)\left(\begin{array}{c}
    e^{x\xi+\Delta_R}e^{i k y
} \\
   e^{-x\xi-\Delta_R}e^{-i k y }
  \end{array}\right).
\end{eqnarray}
 Then we can calculate the density node as the
value of $x$ and $y$ such that $\langle
\psi^\dag(x,y)|\psi(x,y)\rangle=0$
\begin{equation}\label{node}
  x(\varepsilon_F)=-\frac{\Delta_R(\varepsilon_F)}{\xi(\varepsilon_F)}
\qquad y_n(\varepsilon_F)=-\frac{(2n+1)\pi  }{2k(\varepsilon_F)}.
\end{equation}
We can also get the current
\begin{eqnarray}\label{j}
 j_x(x,y)&=&4 f^2(x) \xi \sin(2ky) \nonumber \\
 j_y(x,y)&=&-4 f^2(x) k_0 \sinh(2x\xi+2\Delta_R)
\end{eqnarray}
and the spin
\begin{eqnarray}\label{j1}
 \langle\sigma_x(x,y)\rangle&=& f^2(x) \sinh(2x\xi+2\Delta_R) \nonumber \\
 \langle\sigma_y(x,y)\rangle&=& f^2(x)  \sin(2ky).
\end{eqnarray}
From these formulae we can  obtain the current vector field (which
shows antivortices near the nodes of the wavefunction Fig. (11)
and Fig.(12)) and the spin texture with its very peculiar
topology.


\bibliographystyle{prsty} 
\bibliography{}

\end{document}